\documentstyle[prl,aps,twocolumn,epsfig]{revtex}
\draft
\begin{document}
\bibliographystyle{prsty}

\wideabs{
\title{Boosting Sonoluminescence}
\author{Joachim Holzfuss$^1$, Matthias R\"uggeberg$^1$, Robert Mettin$^{1,2}$}
\address{$^1$Institut f\"ur Angewandte Physik, 
TU Darmstadt, Schlo\ss gartenstr. 7,
64289 Darmstadt, Germany\\
$^2$Drittes Physikalisches Institut, B\"urgerstr. 42-44, 
37073 G\"ottingen, Germany}

\date {Accepted for publication in Phys. Rev. Lett.)\\
(Copyright 1998 by The American Physical Society}
\maketitle

\begin{abstract}
Single bubble sonoluminescence has been experimentally
produced through a novel approach of optimized sound excitation.
A driving consisting of a first and second harmonic with selected
amplitudes and relative phase results in an increase of light emission
compared to sinusoidal driving. We achieved a raise of the maximum
photo current of up to 300\% with the two-mode sound signal. Numerical
simulations of multimode excitation of a single bubble are compared to
this result.
\end {abstract}
}
By focusing ultrasonic waves of high intensity into a liquid, thousands
of tiny bubbles appear. This process of breakup of the liquid is called
acoustic cavitation. The bubbles begin to form a fractal structure that
is dynamically changing in time. They also emit a loud chaotic sound
because of their forced nonlinear oscillations in the sound field
\cite{LauterbornHofu}. The
large mechanical forces on objects brought into contact with the
bubbles enable the usage of cavitation in cleaning, particle
destruction and chemistry.  Marinesco and Trillat \cite{marinesco}
found that a photo plate in water could be fogged
by ultrasound. This multi-bubble sonoluminescence (MBSL) has been
analyzed by many researchers, and a great amount of knowledge
has been gained \cite{walton-young}. The discovery by Gaitan
\cite{felipe} that it is possible to drive a single stable bubble in a
regime, where it emits light pulses of picosecond duration
\cite{putterman1,gompf}, called single bubble sonoluminescence (SBSL),
has been encouraging scientists to explore the phenomenon and the associated
effects with a multitude of experiments, theories and simulations. 
The experimental results show picosecond synchronicity
\cite{barber}, quasiperiodic and chaotic variability of inter-pulse
times \cite{holt94,holzfuss93}, a black body spectrum \cite{hiller}
and mass transport stability \cite{holt96}. The theories to explain the
source of SBSL range from hot spot, bremsstrahlung \cite{WuRoberts}, 
collision induced radiation \cite{fromhold}, and corona discharges 
\cite{lepoint} to non-classical light\cite {schwingereberlein}.  
Numerical simulations have been focusing
on the bubble dynamics, behavior of the gas content
\cite{WuRoberts,moss96}, properties in magnetic fields \cite{Chou} and the
stability of the bubble\cite{brenner}.
However,
the final answer concerning the nature of SBSL still remains open.

The amount of energy concentration from low energy acoustic sound waves
 to 3\ eV photons
\cite{putterman1,barberpr} raises the question, whether the effect can be up-scaled.
In this paper, we report on experimental enhancement of SBSL light
production by a bimodal excitation of the bubble
oscillation. The experiment follows an idea stated in
\cite{daga95}. We also present numerical simulations of multimode
sound driving that reveal how multiharmonic excitation can adapt to the
highly nonlinear bubble oscillation in the sense of a strong collapse.

The experimental setup is as follows: An air bubble is trapped in a
water filled cell consisting of two piezoceramic cylinders connected
via a glass tube \cite{holt90}.  The levitation cell (``Crum cell'')
\cite{crumcell} is standing upright with a glass plate
covering the lower end of the cell. The upper end remains open. A
video camera pointing from the side allows for online monitoring of the
experiment. The experiments were done with distilled and
degassed water
at room temperature and an ambient pressure of 1\ atm. The bimodal driving 
signal $P_e(t)=P_1\cos(2\pi f t)+P_2cos(2\pi2 f t+\phi)$
is produced by synchronized sine wave generators that allow
to fix the amplitudes $P_1$, $P_2$, and the relative phase $\phi$.
  
Using a multifrequency driving signal however is complicated by two
facts. First, the transducers have a complex transfer function; 
second, the standing wave conditions at each frequency in the cell have
to be obeyed \cite{resonances}. 
Therefore, multifrequency driving results in space
dependent
phases and amplitude relations and thus in an effective sound signal
$P_a(r,z,t)$ (with cylinder coordinates $r$,$z$ of the levitation cell).
To measure the amplitudes and relative phase that actually appear at
the bubble position, a small hydrophone is used.
The correct position is adjusted by first focusing the camera on the
bubble and then inserting the 
hydrophone at the bubble site. The driving signal is digitally recorded
and phases and amplitudes are recovered via a Fourier transform. The light
flashes emitted at collapse are measured with a photomultiplier.

Levitating small oscillating bubbles of volume $V(t)$ in nonzero gravity
is possible through the
interaction with the driving sound field
$P_a(r,z,t)$ which depends on space and time.
The time averaged primary Bjerknes force
\cite{crum75}
\begin{equation}
F_B=-\langle V(t) \nabla P_a(r,z,t) \rangle 
\label{PrimaryBjerknes}
\end {equation}
can
overcome the buoyancy force and attract the bubble to a fixed position
in space. Weakly sinusoidally driven bubbles of equilibrium
radius $R_0$ are trapped near a
pressure antinode if they are driven below their linear resonance (Minnaert) frequency 
$f_M={(2\pi R_0)^{-1}\sqrt{3\kappa p_0/\rho}} \approx 3/R_0$[Hz]
for the experimental conditions used here 
(with polytropic exponent
$\kappa$, ambient pressure $p_0$, and liquid density $\rho$)
\cite{minnaert}. However, the situation is more complicated for strongly
driven bubbles \cite{achatov2} and also for multi-modal
excitation, where 
the standing wave pattern in the
resonator, the Bjerknes forces and thus the bubble position are changed
by a variation of the sound signal parameters $P_1$, $P_2$, and $\phi$.
The bubble
oscillation responds to the sound signal at the trapping
site. 

\begin{figure}[htb]
\epsfig{file=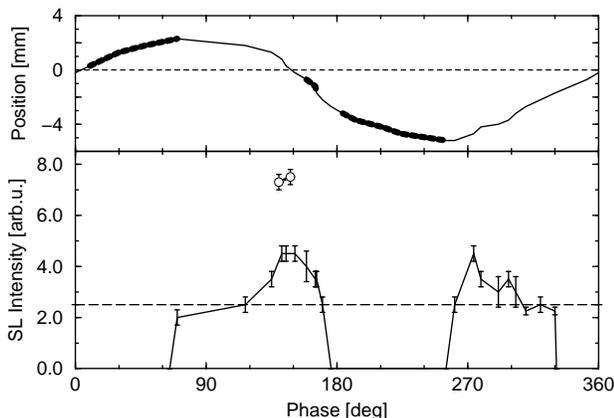,width=5.5cm,angle=-90}
\caption{Bubble response for two-mode driving
as a function of the phase
difference (in degrees) between the driving sinusoidal signal and its second
harmonic. Upper: Vertical position of the bubble. The thick dotted lines denote unstable
bubble behaviour. 
Lower: Photo current. 
The open circles show the maximum SL intensity achieved.
The dashed line is the maximal photo current for
pure sine wave driving.}
\label{explight}
\end{figure}

In the experiment, we proceeded in the following way:
For fixed drive amplitudes $P_1$ and $P_2$, a bubble is injected
into the fluid with a syringe. Once the bubble fixes itself spatially 
at a stable position,
where the Bjerknes force equals the buoyancy force, 
the phase difference between the  locked sine wave
generators, one operating at $f=23.4~{\rm kHz}$ and
the other at 
$2f$, is sequentially increased while the sonoluminescence (SL) intensity and the
bubble itself are monitored. 
Fig.\ \ref{explight} (lower) shows the SL intensity as a
function of the phase difference for 
$P_1 \approx 1.25$\ bars, $P_2 \approx 0.3$\ bar.
With increased phase difference 
two maxima appear in the light intensity.
The dashed line indicates the 
maximum achievable SL intensity using single-mode driving.
This value is obtained shortly before and after the two-mode experiment to allow
direct comparison by keeping all other experimental conditions 
unchanged. It is seen that the two-mode driving yields
100\% more SL intensity than
the maximal single-mode driving. 
By further increase of $P_1$ and $P_2$ at selected phases 
an intensity gain of 300\%
can be achieved, as shown by the open circles. Beyond that, the bubble
gets destroyed.

Fig.\ \ref{explight} (upper) reveals,
that with increased
phase difference the bubble traverses vertically through stable and 
(surface) unstable regimes.

Numerical simulations have been carried out using the Gilmore model
\cite{gilmore} which describes the radial motion of a single
bubble.
The model includes the usual components of the Rayleight-Plesset
equation \cite{loeff} like surface tension $\sigma$ and liquid
viscosity $\mu$, and
also the compressibility of the liquid to allow damping of the bubble
motion by the shedding of shock waves.

\begin{eqnarray}
\left( 1 - \frac{\dot{R}}{C} \right)R\ddot{R}&+&\frac{3}{2}\left( 1
-\frac{\dot{R}}{3 C}\right) \dot{R}^2 \nonumber \\
&=& \left( 1+\frac{\dot{R}}{C}\right) H+
\left(1-\frac{\dot{R}}{C}\right)\frac{R}{C}\frac{dH}{dt},
\label{eq:gilmore}
\end{eqnarray}
\vspace {-0.5cm}
\begin{eqnarray}
  H =\int\limits^{p(R)}_{p_{\infty}}\frac{dp}{\rho} \quad, \quad
  \frac{p(R,\dot{R})+B}{p_0+B}={\left\{\frac{\rho}
  {\rho_{0}}\right\}}^n, \nonumber \\
  \nonumber \\
  \left. c \right|_{r=R}= C =\left. \sqrt{\frac{dp}{d\rho}} \right|_{r=R}=
  \sqrt{c_0^2 + (n-1)H}, \nonumber \\
  \nonumber \\
p(R,\dot{R})=\left(p_{0}+\frac{2\sigma}{R_{0}}\right){\left(\frac{R_{0}^3
-a^3}{R^3-a^3}\right)}^{\kappa}
-\frac{2\sigma}{R}-\frac{4\mu}{R}\dot{R}.\nonumber 
\end{eqnarray}
$R$ is the bubble radius, $R_0=5~\mu$m its equilibrium radius, 
and $C$, $\rho$, and $p$ are the speed of sound in
the liquid, its density, and the pressure at the
bubble wall, respectively. $H$ is the enthalpy  of the
liquid. Parameters were set to $c_0$=1500 m/s,
$\rho_0$=998 kg/m$^3$, $p_0$=1 bar, $\kappa$=4/3, $\sigma$=0.0725 N/m,
$\mu$=0.001 Ns/m$^3$, $n$=7, $B$=3000 bars.
$a$=$R_0$/8.54 is a hard-core van der Waals-term \cite{loeff}. The pressure at
infinity includes the multimodal driving pressure:
$p_\infty=p_0+P_e(t)$, $P_e(t)=\sum_{m=1}^M P_m cos(2\pi m f t
+\phi_m)$.

First we calculated the driving sound signal that would lead to the most
violent collapse, indicated by the smallest minimum radius during a
bubble oscillation cycle using the above equation.
The search for suitable pressures $P_m$ and phases $\phi_m$ was
carried out by a heuristical optimization algorithm
\cite{numrecipes} with the boundary condition of 
a constant driving signal power, i.e., $P_e^2=\sum_{m=1}^M P_m^2=const$.
$P_e$ was fixed to 1.3\ bars and the driving
frequency was the same as in the experiment. 
Comparing equal power signals is convenient, because 
the power stays constant upon
phase changes, making it possible to compare numerics and experiments.

\begin{figure}
\epsfig{file=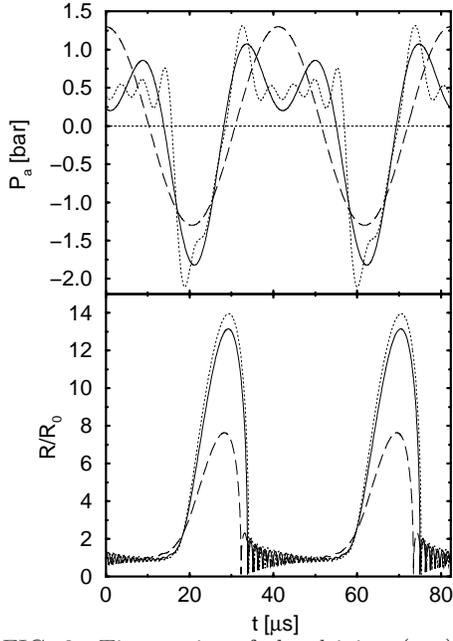,width=6cm}
\caption{Time series of the driving (top) and the radius of calculated
bubble collapses (bottom) for single (dashed) and optimized multimode
driving signals (two-mode: line, eight-mode: dotted).}
\label{calctime}
\end{figure}

\begin{figure}
\epsfig{file=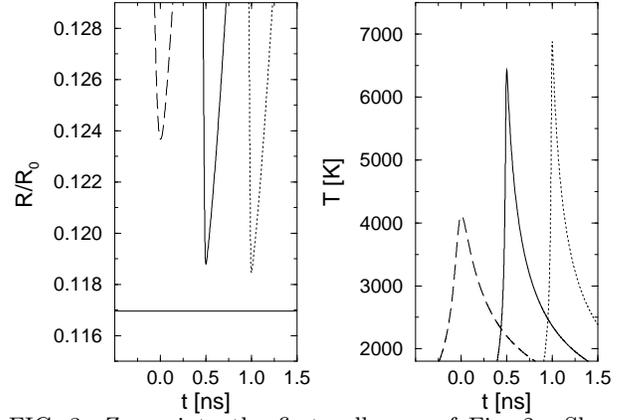,width=8cm}
\caption{Zoom into the first collapses of Fig.\ \ref{calctime}. Shown
are the curves for single-mode (dashed), optimized two-mode (line) and
optimized eight-mode (dotted) driving. The time series of 
the bubble collapses exhibit a decrease in minimum radius (left)
and increase in the adiabatically calculated temperature (right) as a
function of the number of modes in the driving sound.
The minimum radius comes very close to the van der Waals hard core,
shown by the horizontal line in the left graph. The time axis is
shifted so that the collapses take place at 0, 0.5, and 1\ ns, respectively.
}
\label{minmax}
\end{figure}

Fig.\ \ref{calctime} shows the driving pressure and the bubble response
of different driving signals. A strong increase in the maximum radius
can be seen already by adding just the second harmonic to a sine wave.
The numerically computed optimal phase difference is $166.4\ \deg$ and
the individual amplitudes are $P_1=1.026$\ bars and $P_2=0.798$\ bar. The
radius and the adiabatically calculated temperature around the collapses
are shown in Fig.\ \ref{minmax}. It is seen, that the  bubble 
radius at collapse is decreased by a large amount and
is approaching the van der Waals hard core already for the two-mode driving.
Also the maximum temperatures almost double. The higher-mode driving
signals are better adapted to the nonlinear bubble oscillation than
the sine signal: They  show a deeper rarefaction phase before
collapse, followed by a more rapid rise to the compression phase
during collapse.

The calculations for optimal eight-mode driving exhibit only small
additional gain compared to bimodal driving. Because of the
increased  difficulties regarding  
the spatial stability of bubbles in
the resulting complicated sound field, an eight-mode driving may not be
worth being considered experimentally.
Though two-mode driving is an early 
truncation of a series expansion, one sees that already this approximation
shows a trend for a more intense driving of this nonlinear system.

\begin{figure}
\epsfig{file=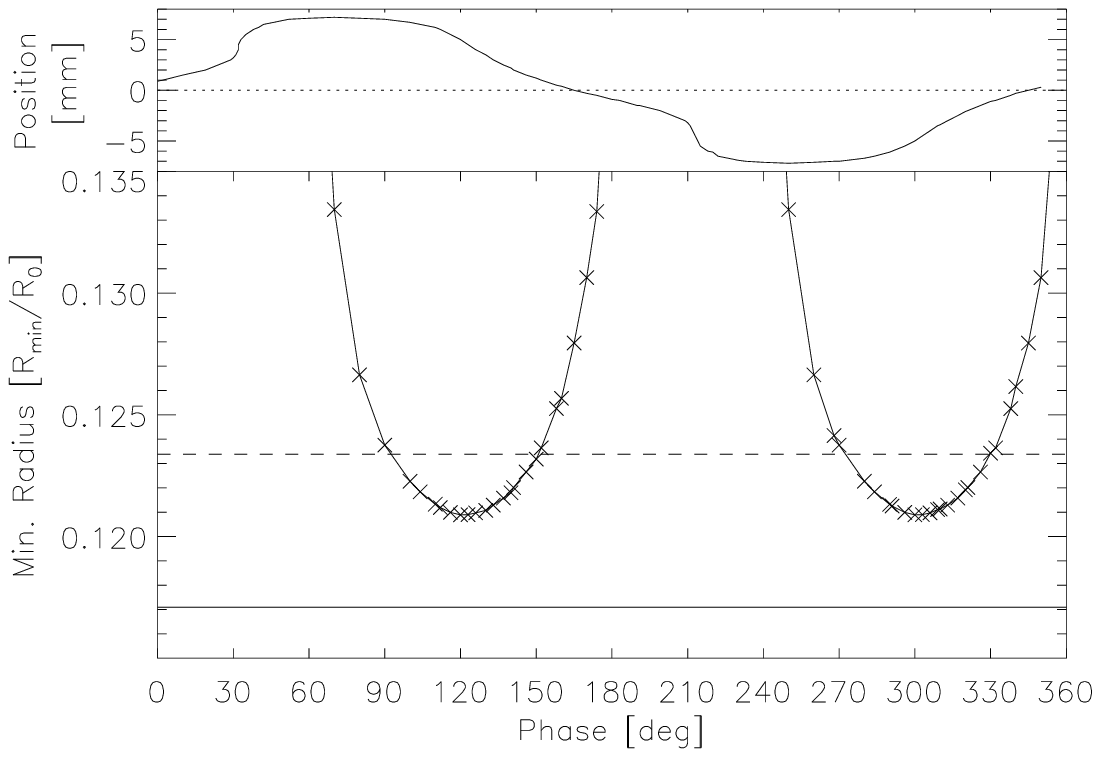,width=8.0cm}
\caption{Numerically calculated vertical bubble position (upper) and 
resulting minimum radius (lower) as a function of the phase difference
for double harmonic driving of a bubble. The dashed line in the lower plot is the 
minimum radius for single frequency driving with the same power. The solid
straight line is the van der Waals hard core.
}
\label{mmbjerk}
\end{figure}

The optimal results are located on a single broad
plateau in parameter space. This is in contrast to the experimental finding
of two maxima. 
To understand the reason of this obvious discrepancy, the bubble model 
[Eq.\ (\ref{eq:gilmore})] has been integrated numerically  along with the
primary Bjerknes [Eq.\ (\ref{PrimaryBjerknes})] and the buoyancy forces to
examine spatial dependencies. 
This is also motivated by the
observation that the vertical position of the bubble is altered when the
phase is changed (Fig.\ \ref{explight} upper). The change of position leads to
different effective excitation amplitudes for $f$ and $2f$ and thus to a more
complex scenario.
The system of equations is integrated using the spatially dependent
driving force
\begin{eqnarray}
P_e(t)=P_1\cos(2\pi f t)&\cos& (k/2~z) \nonumber \\
+P_2&\cos&(2\pi 2 f t+\phi)\cos (2kz-\pi/2) \nonumber
\end{eqnarray} 
($k$ is the acoustic wavenumber 
$2\pi f/c_0$, $P_1=1.25$ bars, $P_2=0.357$ bar, $f=23.4$ kHz). 
The spatial modes are approximately the 
same as the experimental modes, which have been measured with a needle 
hydrophone. The points in vertical $z-$space where the 
Bjerknes force vanishes and the stability
criterion is met represent the position of the bubble. The resulting 
minimum radii are shown in Fig.\ \ref{mmbjerk}. Comparing 
this with the experimental results in Fig.\ \ref{explight}
shows a very close agreement.
\vspace*{-4ex}
The almost sinusoidal variation of the position of the bubble gives
rise to two minima of the minimal radius/phase dependence. 
Each of these minima is smaller
than the one of the single-mode driving. The minima coincide with the
experimental observation of increased SL intensity. The slight asymmetry
in the experiment can be described by the difference in acoustic impedance 
of the glass bottom and the open top of the cell. Also, 
imperfect standing waves may lead to small traveling components in the 
experimental driving. 
Changing the
amplitude ratio of the driving signal closer to unity while keeping the
power constant results in a complex scenario of stable bubble positions and
effective drivings including hysteretic jumps. 


In summary, we have shown that a bimodal sound excitation can enhance
light production of SBSL. Though spatial modes play a crucial role in 
double harmonic driving, it increased the
photo current to a gain of maximally 300\% compared
to sine excitation. We suppose that
multifrequency driving can shift the bubble oscillation to a regime of
strong stable SBSL which is not reachable by pure harmonic
driving. Numerical simulations of an acoustically driven bubble
including Bjerknes and buoyancy forces show that the increased SBSL light intensity
is caused by a larger compression. 
To give quantitative estimates, however, elaborate models have to be considered 
that include gas dynamic equations for
the interior of the 
bubble and thus can model the shedding of a shock wave inside a bubble
\cite{WuRoberts,moss96,moss-science}.

Other methods have been proposed to increase the violence of
bubble collapses. For example, calculations for thermonuclear D-D fusion in
D$_2$O within this context have been done using a
large pressure pulse superimposed on a sine wave\cite{moss96}.
However, whether advanced forcing by higher modes is large
enough to achieve a reasonable neutron production rate is an open
question. Apart from sonoluminescence, the increase of cavitation strength
by means of optimized multiharmonic sound signals \cite{daga95} can also
be of use 
in the context of sonochemistry \cite{suslick} and
related areas, where higher reaction rates could be induced.

The authors wish to thank R.\ G.\ Holt and W. Lauterborn
for stimulating discussions and
the TU Darmstadt for making the research possible.  The work has been
funded through the SFB 185 ``Nichtlineare Dynamik'' of the DFG.


\vspace*{-2ex}
 \begin{thebibliography}{99}
\vspace*{-8ex}
\bibitem{LauterbornHofu}W. Lauterborn, and J. Holzfuss, 
	J. Bif. and Chaos {\bf 1}, 13 (1991); W. Lauterborn,
	and J. Holzfuss,  Phys. Lett. A {\bf 115}, 369 (1986).
\bibitem{marinesco}N. Marinesco and J. J. Trillat,
         Proc. R. Acad. Sci. {\bf 196}, 858 (1933).
\bibitem{walton-young}A. J. Walton and G. T. Reynolds, Adv. Phys.
         {\bf 33}, 595 (1984); F. R. Young, {\it Cavitation} (McGraw-Hill, 
         London, 1989).
\bibitem{felipe}D. F. Gaitan, L. A. Crum, C. C. Church, and R. A. Roy,
         J. Acoust. Soc. Am. {\bf 91}, 3166 (1992).
\bibitem{putterman1}B. P. Barber, S. J. Putterman, Nature {\bf 352},
         318 (1991).
\bibitem{gompf}B. Gompf, R. G\"unther, G. Nick, R. Pecha, and W. Eisenmenger,
	Phys. Rev. Lett. {\bf 79}, 1405 (1997).
\bibitem{barber}B. P. Barber, R. Hiller, K. Arisaka, 
	H. Fetterman, and S. Putterman, 
	J. Acoust. Soc. Am. {\bf 91}, 3061 (1992).
\bibitem{holt94} R. G. Holt, D. F. Gaitan, A. A. Atchley, and J. Holzfuss, 
	 Phys. Rev. Lett. {\bf 72 },  1376 (1994).
\bibitem{holzfuss93}J. Holzfuss, R. G.  Holt, D. F. Gaitan, and A. A. Atchley, 
        in: {\it Fortschritte der Akustik - DAGA 93},
         (DPG GmbH, Bad Honnef,1993), pp. 329-332.
\bibitem{hiller}R. Hiller, S. P. Putterman and B. P. Barber, Phys. Rev. Lett.
	 {\bf 69}, 1182 (1992).
\bibitem{holt96}R. G. Holt and D. F. Gaitan, 
	Phys. Rev. Lett. {\bf 77}, 3791 (1996).
\bibitem{WuRoberts}C. C. Wu and P. H. Roberts, 
	Phys. Rev. Lett. {\bf 70}, 3424 (1993).
\bibitem{fromhold}L. Fromhold, Phys. Rev. Lett. {\bf 73}, 2883 (1994).
\bibitem{lepoint}T. Lepoint, N. Voglet, L. Faille, and 
	F. Mullie, in: {\it Bubble Dynamics and Interface Phenomena},
	edited by: J. R. Blake, J. M. Boulton-Stone, and N. H. Thomas, 
	(Kluwer Academic Publishers, Dordrecht, 1994), pp. 321--333.
\bibitem{schwingereberlein}J. Schwinger, 
	Proc. Natl. Acad. Sci. USA {\bf 89}, 4091 (1992);
	C. Eberlein, Phys. Rev. Lett. {\bf 76}, 3842 (1996).
\bibitem{moss96}W. C. Moss, D. B. Clarke, J. W. White, and D. A. Young, 
	Phys. Lett. A {\bf 211}, 69 (1996)
\bibitem{Chou}T.Chou and E.G. Blackman,
        Phys. Rev. Lett. {\bf 76}, 1549 (1996).  
\bibitem{brenner}M. P. Brenner, D. Lohse, D. Oxtoby, and T. F. Dupont,
         Phys. Rev. Lett. {\bf 76}, 1158 (1996); 
 	 M. P. Brenner, D. Lohse, and T. F. Dupont, Phys. Rev. Lett.
         {\bf 75}, 954 (1995).
\bibitem{barberpr}B. P. Barber, R. A. Hiller, R. L\"offstedt,
	S. J. Putterman, and K. R. Weninger, 
	Phys. Rep. {\bf 281}, 65 (1997).    
\bibitem{daga95}R. Mettin, J. Holzfuss, and W. Lauterborn, in:
	{\it Fort\-schritte der Akustik - DAGA 95},
         (DPG GmbH, Bad Honnef, 1995), pp. 1147-1150.
\bibitem{holt90}R. G. Holt, J. Holzfuss, A. Judt, A. Phillip,  and
         S. Horsburgh, in: {\it Frontiers of Nonlinear Acoustics: 
         Proceedings of the 12th ISNA}, edited by: M.F.
         Hamilton and D.T. Blackstock, (Elsevier Science Publishers
         Ltd., London, 1990), pp. 497-502.
\bibitem{crumcell}L. A. Crum, J. Acoust. Soc. Am. {\bf 68}, 203 (1980). 
\bibitem{resonances}The largest resonances,  Q-factors and amplitudes 
	relative to the main are: 
	(23.4 kHz, 84, 1), (44.4, 21, 0.12), (45.8 kHz, 89, 0.19), 
	(46.8 kHz, 67, 0.17).
\bibitem{crum75}L. A. Crum, J. Acoust. Soc. Am. {\bf 57}, 1363 (1975).

\bibitem{minnaert}M. Minnaert, Phil. Mag. {\bf 16}, 235 (1933).
\bibitem{achatov2}I. Achatov, R. Mettin, C.D. Ohl, U. Parlitz, and
         W. Lauterborn, Phys. Rev E. {\bf 55}, 3747 (1997).
\bibitem{gilmore}F. R. Gilmore,  
         California Institute of Technology Report No. 26-4,
         (1952). 
\bibitem{loeff} R. L\"offstedt, B. P. Barber and S. J, Putterman, Phys. Fluids A, 
	{\bf 5} 2911 (1993).
\bibitem{numrecipes}We used the algorithm {\tt amebsa} from:
         W.H. Press, S.A. Teukolsky, W.T. Vetterling, and B.R.
         Flannery, {\it Numerical Recipes in~C} (Cambridge
         University Press, Cambridge 1992) 2nd~ed.
\bibitem{moss-science}W. C. Moss, D. B. Clarke, and D. A. Young, Science 
	{\bf 276}, 1398 (1997).


\bibitem{suslick} {\it Ultrasound: Its Chemical,
         Physical and Biological Effects} edited by K. S. Suslick,
         (VCH, New York, 1988) 



	

\end{thebibliography}
\end{document}